\documentclass[prl,twocolumn,tightenlines,nofootinbib]{revtex4-1}
\usepackage{bm,dcolumn,amsmath,graphicx,amsfonts,amssymb}

\usepackage{hyperref}
\usepackage{xcolor}
\definecolor{cite}{rgb}{0.,0.,0.9}   
\hypersetup{colorlinks,linkcolor={cite},citecolor={cite},urlcolor={cite}}

\renewcommand{\v}[1]{\ensuremath{\boldsymbol{#1}}}		

\newcommand{\bra}[1]{\ensuremath{\langle #1|}}	
\newcommand{\ket}[1]{\ensuremath{|#1\rangle}}	
\newcommand{\braket}[1]{\ensuremath{\langle #1\rangle}}	

\newcommand{\smallspace}{\rule{0pt}{2.5ex}}

\newcommand{\A}{\ensuremath{\mathcal{A}}} 

\begin{document}

\title{The hyperfine anomaly in heavy atoms and its role \\in precision atomic searches for new physics}

\author{B.\ M.\ Roberts}\email[]{b.roberts@uq.edu.au}
\author{J.\ S.\ M.\ Ginges}\email[]{j.ginges@uq.edu.au}
\affiliation{School of Mathematics and Physics, The University of Queensland, Brisbane QLD 4072, Australia}
\date{\today}

\begin{abstract}\noindent
We report on our calculations of differential hyperfine anomalies in the nuclear single-particle model for a number of atoms and ions of interest for studies of fundamental symmetries violations. 
Comparison with available experimental data allows one to discriminate between different nuclear magnetization models, and this data supports the use of the nuclear single-particle model over the commonly-used uniform ball model.
Accurate modelling of the nuclear magnetization distribution is important for testing atomic theory through hyperfine comparisons. 
The magnetization distribution must be adequately understood and modelled, with uncertainties well under the atomic theory uncertainty, for hyperfine comparisons to be meaningful. 
This has not been the case for a number of atoms of particular interest for precision studies, including Cs.
Our work demonstrates the validity of the nuclear single-particle model for Cs, and this has implications for the theory analysis of atomic parity violation in this atom.
\end{abstract}

\maketitle



Investigations of atomic parity violation provide some of the most constraining low-energy tests of electroweak theory~\cite{GingesRev2004,RobertsReview2015,AtomicReview2017,PDG2020}.
These investigations require exquisitely precise measurements of parity-violating transition amplitudes~\cite{Wieman1997,Wieman1999,TohBeta2019}, and equally precise atomic structure calculations~\cite{GingesCs2002,FlambaumQED2005,Porsev2009,*Porsev2010,OurCsPNC2012} for their interpretation.
Similarly, measurements of time-reversal-violating electric dipole moments (EDMs) in atoms and molecules require atomic and molecular structure theory for their interpretation in terms of fundamental charge-parity (CP) violating parameters~\cite{Pospelov2005,Engel2013,Cairncross2019,Chupp2019}.
While such EDMs have eluded detection to date, the experimental programs are ramping up and their measurements clamping down on the size of these EDMs~\cite{Regan2002,Hudson2011,Parker2015,Graner2016,JILA-EDM2017,NLeEDM2018,ACME2018,ACME2020,Grasdijk2020}.
The implications for new CP-violating models are profound, demanding increasingly accurate theory for meaningful constraints and in anticipation of non-zero measurements.

The magnetic hyperfine structure, which arises due to the interaction of atomic electrons with the nuclear magnetic moment, plays an important role in
precision studies of violations of fundamental symmetries.
The testing and further development of atomic theory depends on comparisons between calculated and measured quantities that probe the atomic wavefunctions across all length scales of the atom.
The quantities used for benchmarking include binding energies, electric dipole matrix elements, and hyperfine structure constants. It is the latter that allows unique access to the quality of the wavefunctions in the nuclear region,
where the parity-violating and EDM interactions take place; see, e.g., Ref.~\cite{GingesRev2004}.

Hyperfine structure calculations depend on the modelling of nuclear structure effects.
In particular, they are sensitive to the distribution of the nuclear magnetic moment across the nucleus, the so-called Bohr-Weisskopf effect~\cite{Bohr1950,Bohr1951}.
It has been recognised only recently~\cite{Ginges2017,Ginges2018,RobertsFr2020} that for a number of heavy atoms of interest this dependence is much stronger than has been assumed.
The effect is so large that in some cases the hyperfine splitting shifts by more than the claimed atomic theory uncertainty when switching from one nuclear magnetization model to another.
Indeed, for Cs and Fr -- of particular interest in atomic parity violation studies, and where the claimed atomic structure uncertainty has reached 0.5\% or better -- the hyperfine splittings change by as much as \,0.5\% and 1.5\%, respectively, when a simple nuclear single-particle model is used in place of the widely-adopted uniform distribution.
The ability to test the validity of these models is therefore critically important to the field.

The most precise atomic parity violation measurement has been performed for Cs~\cite{Wieman1997}, and there are new experiments underway in Cs~\cite{TohBeta2019} and Fr~\cite{Aubin2013,Tandecki2013} and interest in studying Ba$^+$, Ra$^+$, and Rb~\cite{Fortson1993,DzubaPNCsd2001,Koerber2003,Pal2009,Mandal2010,Versolato2011,OurRb2012,RobertsActinides2013,*Robertssd2014,NunezPortela2014,Fan2019}.
Measurements of atomic parity violation across a chain of Yb isotopes have recently been performed~\cite{Antypas2018a}, and while the considered ratios of measured values do not rely on atomic structure for their interpretation, they strongly depend on the neutron distribution.
Systems under recent and ongoing experimental investigation for detection of EDMs include the paramagnetic atoms Tl~\cite{Regan2002},   
Fr~\cite{Harada2016}, and molecules YbF~\cite{Hudson2011}, BaF~\cite{NLeEDM2018}, 
and the diamagnetic systems Hg~\cite{Graner2016}, Ra~\cite{Parker2015}, and TlF~\cite{Grasdijk2020}.

In this work, we calculate the Bohr-Weisskopf effect and differential hyperfine anomalies for systems of interest for precision atomic studies that may be treated as single-valence-electron atoms or ions and for which there is experimental data to compare -- Rb, Cs, Ba$^+$, Yb$^+$, Hg$^+$, and Tl.
We studied these effects in Fr isotopes in our recent paper in which improved nuclear magnetic moments were deduced~\cite{RobertsFr2020}, and they were studied more recently in thallium isotopes \cite{Prosnyak2021}.
The differential hyperfine anomaly gives the difference in the hyperfine structure for different isotopes of the same atom that arises due to finite nuclear size effects.
We show that available experimental data
allows one to distinguish between the different nuclear magnetization models.
This data supports the use of the nuclear single-particle model~\cite{LeBellac1963,Shabaev1994,Shabaev1995,volotka08a,Shabaev1997}, rather than the uniformly magnetized ball, for modelling the Bohr-Weisskopf effect.


The relativistic operator for the electron interaction with the nuclear magnetic moment is
\begin{equation}
\label{eq:h-hfs}
h_{\rm hfs}
= \alpha\, {\v{\mu}\cdot (\v{r}\times \v{\alpha})} \, F(r) /r^{3} ,
\end{equation}
where $\v{\alpha}$ is a Dirac matrix, $\v{\mu}=\mu \v{I}/I $ is the nuclear magnetic moment, $\v{I}$ is the nuclear spin, and
$F(r)$ describes the nuclear magnetization distribution [$F(r)=1$ for a pointlike nucleus], and $\alpha\approx1/137$ is the fine-structure constant
(we use atomic units $\hbar$\,=\,$|e|$\,=\,$m_e$\,=\,1, $c$\,=\,$1/\alpha$).
The expectation value of the operator (\ref{eq:h-hfs}) may be expressed as $\A\braket{\v{I}\cdot\v{J}}$, where
$\v{J}$ is the electron angular momentum, and $\A$ is the magnetic dipole hyperfine constant.

The Bohr-Weisskopf (BW) effect arises from the finite nuclear magnetization distribution and gives a significant contribution to the hyperfine structure~\cite{Bohr1950}.
For heavy atoms, it has been standard to model the  nucleus as a ball of uniform magnetization, such that
\begin{equation}\label{eq:Fball}
F_{\rm Ball}(r) =
\begin{cases}
(r/r_m)^3 & r < r_{m} \\
1 &  r \geq r_{m}
\end{cases},
\end{equation}
with the nuclear magnetic radius typically taken as $r_{m}$\,=\,$\sqrt{5/3}\,r_{\rm rms}$, where $r_{\rm rms}$ is the root-mean-square (rms) charge radius.

A more sophisticated modelling of the magnetization distribution, that takes into account the nuclear angular momenta and configuration, may be given by the simple nuclear single-particle (SP) model~\cite{LeBellac1963,Shabaev1994,volotka08a}. For odd isotopes, we take the distribution as presented in Ref.~\cite{volotka08a},
\begin{equation}\label{eq:F-SP}
F_I(r) = F_{\rm Ball}(r) \big[ 1- \delta F_I \, \ln(r/r_m) \, \Theta(r_m-r) \big],
\end{equation}
where $\Theta$ is the Heaviside step function and
\begin{equation}\label{eq:dFI}
\delta F_I =
\begin{cases}
\dfrac{3(2I-1)}{8(I+1)}\, \dfrac{4(I+1) g_L - g_S}{g_I I}  & I=L+1/2 \\
\dfrac{3(2I+3)}{8(I+1)}\, \dfrac{4I g_L + g_S}{g_I I}   & I=L-1/2
\end{cases}.
\end{equation}
Here, $I$, $L$, and $S$ are respectively the total, orbital, and spin angular momenta for the unpaired nucleon~\cite{volotka08a},
$g_L = 1(0)$ for a proton(neutron), and $g_I = \mu/(\mu_N I)$ is the nuclear $g$-factor with $\mu_N$ the nuclear magneton.
The spin $g$-factor, $g_S$, is chosen so that the experimental value for $g_I$ is reproduced using the Land\'e $g$-factor expression. Formulae (\ref{eq:F-SP}),\,(\ref{eq:dFI}) are found by taking the radial part of the probability density of the nucleon to be constant across the nucleus.
The model may be improved, e.g., by finding the nucleon wavefunction in a Woods-Saxon potential and including the spin-orbit interaction~\cite{Shabaev1997}.
The effect of accounting for these has been shown to be small ($\lesssim 10\%$) for the Bohr-Weisskopf effect in $^{87}$Rb, $^{133}$Cs, and $^{211}$Fr \cite{Ginges2017}, as well as in isotopes of Tl \cite{Prosnyak2021}, and larger in $^{135}$Ba$^{+}$ and $^{225}$Ra$^{+}$~\cite{Ginges2017}.
The single-particle model may be extended in a simple way to describe the magnetization distribution of doubly-odd (odd proton, odd neutron) isotopes~\cite{Buttgenbach1984,Shabaev1995,RobertsFr2020}.

The BW effect may be parameterized as~\cite{Persson2013}
\begin{equation}\label{eq:A0eps}
\A = \A_0(1+\epsilon),
\end{equation}
where $\A_0$ is the hyperfine constant with a pointlike  magnetization distribution ($F$\,=\,1).
Here, $\A_0$ includes the Breit-Rosenthal correction, $\delta$, due to the finite nuclear charge distribution, which is taken into account by solving the electron wavefunctions in the field of a finite nucleus (we use a Fermi distribution with rms charge radii from Ref.~\cite{Angeli2013}).
This may be expressed as $\A_0 = \A_{00}(1+\delta)$, where $\A_{00}$ is the hyperfine constant with pointlike nuclear magnetic and charge distributions.
Since the nuclear charge distribution is known with relatively high accuracy, errors associated with the Breit-Rosenthal correction are typically negligible~\cite{Rosenberg1972,RobertsFr2020,Heggset2020}.
Note that radiative quantum electrodynamics (QED) corrections contribute to the hyperfine structure with comparable size to $\epsilon$ \cite{Shabaev1997,Sapirstein2003b,Ginges2017}, though they are largely independent of the isotope and therefore mostly cancel in the differential hyperfine anomaly considered below. Therefore, we don't consider QED contributions further.

\begin{table*}
\caption{Bohr-Weisskopf corrections, $\epsilon$, and hyperfine anomalies, $^1\Delta^2$, calculated in the ball and single-particle (SP) nuclear magnetization models for the lowest states of several atoms of interest, and comparison with experimental differential anomalies.
$A$ is the atomic mass number for the isotope, $I^\pi$ is the nuclear spin and parity.}
\label{tab:anomaly}
\begin{ruledtabular}
\begin{tabular}{lllc dd lc dd dD{.}{.}{1.3} D{.}{.}{1.8}}
&&\multicolumn{4}{c}{Isotope 1} &
\multicolumn{4}{c}{Isotope 2}  &
\multicolumn{3}{c}{Differential anomaly $^1\Delta^2$ (\%)} \\
\cline{3-6}\cline{7-10}\cline{11-13}
\smallspace
& &
\multicolumn{1}{c}{$A$} &
\multicolumn{1}{c}{$I^{\pi}$} &
\multicolumn{1}{c}{$\epsilon_{\rm Ball}$\,(\%)} &
\multicolumn{1}{c}{$\epsilon_{\rm SP}$\,(\%)} &
\multicolumn{1}{c}{$A$} &
\multicolumn{1}{c}{$I^{\pi}$} &
\multicolumn{1}{c}{$\epsilon_{\rm Ball}$\,(\%)} &
\multicolumn{1}{c}{$\epsilon_{\rm SP}$\,(\%)} &
\multicolumn{1}{c}{Ball} &
\multicolumn{1}{c}{SP} &
\multicolumn{1}{c}{Expt.~\cite{Persson2013}} \\
\hline
\smallspace
$_{37}$Rb &$5s_{1/2}$ & 85 &5/2$^-$& -0.306 & 0.044  & 87 &3/2$^-$& -0.306 & -0.278 & -0.001  & 0.323  & 0.35142(30) \\
   &&     & &       &        & 86  &2$^-$& -0.306 & -0.139 & 0.000  & 0.183  & 0.17(9)     \\
\smallspace
$_{47}$Ag &$5s_{1/2}$ &  107 & 1/2$^-$& -0.497& -4.20  & 103 &7/2$^+$& -0.493 & -0.347 & -0.018   & -3.88  & -3.4(17)    \\
&&&&& & 109 &1/2$^-$& -0.498 & -3.78 & 0.007  & -0.431  &-0.41274(29) \\
\smallspace
$_{55}$Cs &$6s_{1/2}$& 133 & 7/2$^+$& -0.716 & -0.209 & 131 &5/2$^+$& -0.716 & -0.596 & -0.001 & 0.389  & 0.45(5)\tablenotemark[1]     \\
   &&     &         &&        & 135 &7/2$^+$& -0.716 & -0.247 &0.002   & 0.039 & 0.037(9)\tablenotemark[2]      \\
   &&     &        &&        & 134 &4$^+$& -0.716 & -0.371 & 0.000   & 0.163  & 0.169(30)     \\
\smallspace
$_{56}$Ba$^+$ &$6s_{1/2}$ & 135 &3/2$^+$& -0.747 &	-1.03 & 137	&3/2$^+$&	-0.747 &	-1.03&  0.001&	0.001&-0.191(5)\\
$_{70}$Yb$^+$ &$6s_{1/2}$ & 171 &1/2$^-$& -1.37 & -2.41 & 173 &5/2$^-$& -1.38 & -1.79 & 0.014 & -0.618  & -0.425(5)     \\
$_{79}$Au &$6s_{1/2}$ &  197 & 3/2$^+$& -1.97& 15.5 & 199 & 3/2$^+$& -1.97 & 7.47 & 0.013   & 7.48  & 3.64(29)   \\
$_{80}$Hg$^+$& $6s_{1/2}$ & 199 &1/2$^-$&	-2.04&	 -3.57 &	201 &3/2$^-$&	-2.05&	-2.20&  0.016&	-1.39&	-0.16257(5)\\
$_{81}$Tl& $7s_{1/2}$ & 203 &1/2$^+$& -2.13 & -2.13 & 205 &1/2$^+$& -2.13 & -2.13 & 0.015   & 0.015 & 0.0294(81)\tablenotemark[3]  \\
$_{81}$Tl&$6p_{1/2}$ &    203     &1/2$^+$& -0.780 & -0.780 &   205  &1/2$^+$& -0.781 & -0.781 & 0.005   & 0.005 &  0.01035(15)\\
\end{tabular}
\tablenotetext[1]{Ref.~\cite{Worley1965}, Ref.~$^{\rm b}$\cite{Stroke1957}, Ref.~$^{\rm c}$\cite{Chen2012}.}
\end{ruledtabular}
\end{table*}

We calculate $\A_0$ using the relativistic Hartree-Fock method, including the important core-polarization contribution by means of the time-dependent Hartree-Fock (TDHF) method~\cite{DzubaHFS1984,Dzuba1987jpbRPA}, equivalent to the random phase approximation with exchange (RPA).
We consider atoms with a single valence electron above a closed-shell core, for which the valence wavefunction is found in the Hartree-Fock potential due to the ($N-1$) core electrons ($N=Z$ for neutral atoms).
The set of TDHF equations,
\begin{equation}\label{eq:tdhf}
(H - \varepsilon_c)\delta\psi_c = -(h_{\rm hfs} +\delta V-\delta\varepsilon_c)\psi_c\,,
\end{equation}
is then solved self-consistently for each electron in the core.
Here, $H$, $\psi_c$, and $\epsilon_c$ are the relativistic Hartree-Fock Hamiltonian, core electron orbitals, and core electron binding energies, respectively, and $\delta\psi_c$ and $\delta\varepsilon_c$ are hyperfine-induced corrections for core orbitals and energies.
The resulting hyperfine-induced correction to the Hartree-Fock potential is given by $\delta V$.
Since $h_{\rm hfs}$ can mix states with different angular momenta, $\delta\psi_c$ is not an angular momentum eigenstate and contains contributions from states with
$j = j_c, \, j_c\pm1$ ($j_c$ is the angular momentum of single-electron state $c$).
Then, matrix elements for valence states, $v$, are calculated as $\bra{v}h_{\rm hfs}+\delta V\ket{v}$, which includes the core-polarization effects to all-orders in the Coulomb interaction~\cite{Dzuba1987jpbRPA}.
Correlation corrections to hyperfine structure were studied recently by us in detail~\cite{Grunefeld2019}, and those beyond core polarization were shown not to be important for the relative BW effect (see also Ref.~\cite{Ginges2017,Ginges2018,Konovalova2020}).
The insensitivity of the relative BW effect to correlations is due to the short-range nature of the effect, with account of correlations affecting the normalization of the wave functions which largely factors out in the relative correction.


The differential hyperfine anomaly, ${}^1\Delta^2$, is defined via the ratio of the hyperfine constants for different isotopes of the same atom (see, e.g.,~\cite{Persson2013}):
\begin{equation}\label{eq:anomaly}
\frac{\A^{(1)}}{\A^{(2)}} = \frac{g_I^{(1)}}{g_I^{(2)}}\left(1+{}^1\Delta^2\right).
\end{equation}
This quantity, which is a measure of the deviation of the hyperfine structure from the case of a pointlike nucleus, may be found with high accuracy from experiment, provided the nuclear magnetic moments are known well and determined independently of the hyperfine measurements~\cite{Persson2013}.
As for the theoretical determination of $^1\Delta^2$, the correlation corrections beyond RPA that contribute to the hyperfine constants $\A^{(1)}$ and $\A^{(2)}$ cancel in the ratio~\cite{Grunefeld2019,Konovalova2020}, making the electronic structure calculations robust at the level of RPA and of high accuracy. 
Note that there is a strong cancellation of the Breit-Rosenthal corrections, $\delta^{(1)}-\delta^{(2)}$, in the differential anomaly, and typically the differential hyperfine anomaly is strongly dominated by the differential Bohr-Weisskopf effect~\cite{Persson2013},
\begin{equation}\label{eq:anomaly2}
{}^1\Delta^2  \approx\epsilon^{(1)} - \epsilon^{(2)}.
\end{equation}
Therefore, comparison of calculated and measured hyperfine anomalies presents a powerful test of the validity of nuclear magnetization models.

In Table~\ref{tab:anomaly} we present our results for the BW effects and differential hyperfine anomalies obtained using the ball (\ref{eq:Fball}) and single-particle (\ref{eq:F-SP}) models.
The numerical accuracy for the BW calculations is better than 1\%, well below the model uncertainty.
We present results for the lowest states of systems of interest for atomic parity violation and electric dipole moment studies, and we also present results for Ag and Au, which may be treated as single valence electron systems and for which the BW effects and hyperfine anomalies are particularly large. 
The anomaly is calculated using Eq.~(\ref{eq:anomaly}) rather than Eq.~(\ref{eq:anomaly2}), which means that the small differential Breit-Rosenthal effect is included.
The calculated values for $^1\Delta^2$ are compared against available experimental data~\cite{Persson2013}.
Note that the uncertainties in the measured values are dominated by uncertainties in the nuclear magnetic moments~\cite{Stone2005}.

We draw attention to several points.
Firstly, the BW correction is a significant effect, typically entering at the level of several $0.1\%$ to several $1\%$ for the considered systems. 
For Ag and Au the effect is even larger, contributing at around $10\%$ for Au.
Secondly, the ball and single-particle models often lead to substantially different BW effects. 
For $^{85}$Rb and $^{133}$Cs, the difference is as large as $0.4\%$ and $0.5\%$, respectively, matching or exceeding the atomic structure theory uncertainty~\cite{Ginges2017,RobertsFr2020}. 
For $^{171}$Yb and $^{199}$Hg, this difference is 1.0\% and 1.5\%. On the other hand, for $^{203}$Tl and $^{205}$Tl, the two models give the same BW effect due to their nuclear spin-parity being $1/2^{+}$.
Finally, from a comparison of the calculated and measured hyperfine anomalies in Table~\ref{tab:anomaly}, it is seen that the nuclear SP model leads to substantially better agreement with experiment for the majority of cases. 
The agreement is particularly good for isotopes of Rb and Cs.
For example, for Cs the SP model gives $^{133}\Delta^{131}=0.389\%$, $^{133}\Delta^{134}=0.163\%$, and  $^{133}\Delta^{135}=0.039\%$ in excellent agreement with the measured data $0.45(5)\%$, $0.169(30)\%$, and $0.037(9)\%$, respectively.
We note that the single-particle model also works very well for Ag and it works reasonably well for Yb$^+$ and Au, including reproducing the (atypical) sign of the effect for the latter.
Our calculations for Au are in good agreement with previous calculations~\cite{Bout1967,Ekstrom1980}, though we use a simplified model (see also \cite{Stroke1961,Barzakh2020,Demidov2020}).
In the ball model, the only difference in the BW effect between isotopes comes from changing the nuclear radius, similarly to the differential Breit-Rosenthal effect, so the calculated anomaly is always small and the model generally cannot produce the observed anomalies.

Recently, we considered the case of Fr in detail, and we demonstrated using ``double'' differential hyperfine anomalies \cite{Persson1998} that the single-particle model works very well for both odd and doubly-odd isotopes between 207--213~\cite{RobertsFr2020} (see also Refs.~\cite{Grossman1999,Zhang2015,Konovalova2018}).
This allowed us to extract nuclear magnetic moments for these isotopes with significantly higher precision than was previously possible.
The BW effect is particularly large for these Fr isotopes ($\sim$\,1--2\%), and it differs between the ball and SP models by nearly 1.5\% for the odd isotopes.
This difference is the main reason for the large deviation (up to $2\%$) of the deduced moments found in Ref.~\cite{RobertsFr2020} from previous best determinations.
In the same work, the validity of the nuclear SP model in the neighborhood of Fr was supported by the BW effect extracted directly from hydrogenlike $^{207}$Pb and $^{209}$Bi: the calculated and experimental BW results for $^{207}$Pb$^{81+}$ are $-3.55\%$ and $-3.85(4)\%$, respectively, and for $^{209}$Bi$^{82+}$ are $-1.07\%$ and $-1.03(5)\%$ (see \cite{RobertsFr2020} and references therein).

It is worth discussing the case of the Bohr-Weisskopf effect in $^{225}$Ra$^+$, which was calculated in Ref.~\cite{Ginges2017}. 
Generally, the size of atomic structure uncertainties precludes the direct extraction of the BW effect from hyperfine comparisons in many-electron systems. 
For this system, however, the effect is so large that direct extraction is possible. 
This was done in Ref.~\cite{Skripnikov2020} where the value $\epsilon=-4.7\%$ was obtained, which may be compared to the simple SP (and ball) result $\epsilon = -2.8\%$ and to the more sophisticated SP result with the nucleon wave function found in the Woods-Saxon potential and with spin-orbit interaction included, $\epsilon = -4.3\%$ \cite{Ginges2017}.

The simple nuclear single-particle model does not always work well.
This may be seen from Table~\ref{tab:anomaly} for Ba$^+$, Hg$^+$, and Tl.
For the considered isotopes of Hg$^+$, the SP model produces a differential hyperfine anomaly that is significantly larger than the observed value.
For isotopes of Ba$^+$ and Tl, the nuclear states are the same, and the SP model produces very similar BW effects which cancel strongly in the anomaly (\ref{eq:anomaly2}). 
In this case, the neglected nuclear many-body contributions will be more important for the differential anomaly than for the BW effect. 
Another reason for the discrepancy that appears in the hyperfine anomaly may arise due to the magnetic radius.
In our calculations we have taken the magnetic rms radius to be the same as the charge rms radius, though there is no reason for them to be the same.
Indeed, there are indications that the magnetic radii for $^{203}$Tl and $^{205}$Tl are different from one another and from the charge radius \cite{Beiersdorfer2001,Prosnyak2020,RobRanGin2020}. 
It was shown very recently that the nuclear single-particle model outperforms the ball model for several Tl isotopes with different nuclear states~\cite{Prosnyak2021}, 
and the BW effects extracted from experiments with $^{203,205}$Tl$^{80+}$ are in good agreement with nuclear SP calculations \cite{Beiersdorfer2001,Shabaev1997}.
For Ba$^+$, we can look to the Cs differential anomalies involving $^{134}$Cs, with the same neutron configuration as $^{135}$Ba$^+$; coincidence of the SP and experimental results (Table~\ref{tab:anomaly}) lends support to the validity of the model, which describes the magnetization distribution of the unpaired neutron.
While the nuclear single-particle model does not always give differential anomalies in good agreement with experiment, it generally performs better than the ball model across the board and is expected to produce more accurate values for the BW effect.


We now consider the implications for atomic parity violation studies.
The dominant parity violating effects in atoms arise due to the exchange of neutral weak bosons between atomic electrons and the nucleus, leading to the mixing of atomic states of opposite parity; see, e.g., Ref.~\cite{GingesRev2004}.
Such interactions are localized on the nucleus, and the theoretical evaluation of the relevant matrix elements, e.g.,
$\bra{s_{1/2}}\hat H_{\rm PV}\ket{p_{1/2}},$
therefore depends on precise knowledge of the wavefunctions in this region.
These matrix elements cannot be directly compared to experiment, and information about the accuracy of the wave functions and the matrix elements is found from a survey of the deviations between theory and experiment for the hyperfine constants of the relevant states and for the combination $\sqrt{\A^s\A^p}$,
which is considered to give a more reliable indication of the accuracy for the off-diagonal matrix elements~\cite{GingesCs2002,GingesRev2004}.

Separating out the Bohr-Weisskopf contribution, the relevant quantity becomes
\begin{equation}
\label{eq:asap}
\sqrt{\A^s \A^p}\approx \sqrt{\A^s_0\, \A^p_0}\, \Big[1+(\epsilon_s + \epsilon_p)/2\Big] \, .
\end{equation}
The claimed accuracy of the most precise atomic parity violation calculations for $^{133}$Cs was based in part on deviations of the hyperfine constants and the quantity (\ref{eq:asap}) from experiment, where the nuclear magnetization distribution was treated as uniformly distributed (ball model). 
Correcting the nuclear magnetization model (to single-particle) leads to a significant change in the hyperfine constants for $s$ states by $+0.5\%$. 
(For $p_{1/2}$ states, the BW effect is an order of magnitude smaller and the hyperfine constants change by only $0.03\%$.) 
Evaluations of the quantity (\ref{eq:asap}) should therefore be shifted by $+0.3\%$, which is large compared to the highest-precision atomic parity violation calculations with claimed uncertainties 0.27\%-0.5\% \cite{GingesCs2002,FlambaumQED2005,Porsev2009,*Porsev2010,OurCsPNC2012}. 
The effect of this correction is an increase in the deviation from experiment for the hyperfine $6s$ state and the associated quantity (\ref{eq:asap}) in Ref.~\cite{Porsev2009,*Porsev2010}, while the results of the work \cite{GingesCs2002} are hardly changed when accounting also for QED contributions, which is consistent with the 0.5\% claimed uncertainty for the atomic parity violation calculation in that work \cite{GingesCs2002,FlambaumQED2005}; see the analysis in Ref.~\cite{Ginges2018}. This illustrates the importance of understanding and controlling the nuclear magnetic structure, for both reliable benchmarking and continued development of precision atomic theory, and for assigning atomic theory uncertainty which has ramifications for constraints on new physics.


In summary, hyperfine comparisons form an important part of the atomic theory analysis in atomic tests of fundamental physics, most notably atomic parity violation and EDM studies in atoms and molecules. 
These comparisons are only reliable if the nuclear magnetization distribution is adequately understood and modelled, with uncertainties well under the atomic theory uncertainty.
This has not been the case for a number of atoms of particular interest for precision studies, including Cs. 
In this work, we point out that sufficient experimental data exists for many isotopes of interest to be able to test nuclear magnetization models using hyperfine anomalies.
From a study of the hyperfine anomalies, we demonstrate that the single-particle model generally outperforms the near universally-used ball model. It is simple enough to include into atomic structure codes without the need for any sophisticated nuclear calculations, and we advocate its use in future studies.
These investigations into hyperfine anomalies open a new window for probing nuclear structure, including the neutron distribution.

\subsection{Acknowledgments}
This work was supported by the Australian Government through an Australian Research Council Future Fellowship, Project No.~FT170100452.

\bibliography{hfs,extra}

\end{document}